\documentclass[aps,pra,twocolumn,showpacs,floatfix]{revtex4}
\usepackage{epsfig}
\usepackage{graphicx}
\usepackage{dcolumn}
\usepackage{amsthm,amsmath}
\usepackage{longtable}

\begin{document}
\title{Magic wavelengths in the alkaline earth ions }
\author{Jasmeet Kaur$^a$, Sukhjit Singh$^a$, Bindiya Arora$^a$\footnote{Email: arorabindiya@gmail.com} 
and B. K. Sahoo $^{b}$\footnote{Email: bijaya@prl.res.in}}
\affiliation{$^a$Department of Physics, Guru Nanak Dev University, Amritsar, Punjab-143005, India}
\affiliation{$^b$Theoretical Physics Division, Physical Research Laboratory, Navrangpura, Ahmedabad-380009, India}
\date{Received date; Accepted date}

\begin{abstract}
We present magic wavelengths for the $nS$ - $nP_{1/2,3/2}$ and $nS$ - $mD_{3/2,5/2}$ transitions, with the respective ground and 
first excited $D$ states principal quantum numbers $n$ and $m$, in the Mg$^+$, Ca$^+$, Sr$^+$ and Ba$^+$ alkaline earth ions for 
linearly polarized lights by plotting dynamic polarizatbilities of the $nS$, $nP_{1/2,3/2}$ and $mD_{3/2,5/2}$ states of the ions.
These dynamic polarizabilities are evaluated by employing a relativistic all-order perturbative method and their accuracies are ratified 
by comparing their static values with the available high precision experimental or other theoretical results. Moreover, some
of the magic wavelengths identified by us in Ca$^+$ concurs with the recent measurements reported in [{\bf Phys. Rev. Lett. 114, 223001 
(2015)}]. Knowledge of these magic wavelengths are propitious to carry out many proposed high precision measurements trapping the above 
ions in the electric fields with the corresponding frequencies. 
\end{abstract}
\maketitle

State-insensitive trapping techniques have lead to tremendous advancements in the manipulation and control of atoms in far detuned 
optical traps. In this approach, the atoms are trapped at the wavelengths (related to frequencies) of an external electric field at 
which the differential light shift of an atomic transition, that is intended to be probed, due to the Stark effects nullify. These 
wavelengths are specially referred to as magic wavelengths ($\lambda_{\rm{magic}}$s)\cite{Katori}. It has been demonstrated earlier ability of 
trapping neutral atoms inside high-$Q$ cavities at $\lambda_{\rm{magic}}$s in the strong coupling regime, 
which is important in the quantum computation and communication schemes \cite{Keever}. This technique is now widely used to 
carry out many high precision measurements by eliminating large systematics due to stray electric fields. Another notable application 
of these wavelengths is to perform clock frequency measurements \cite{Margolis}, especially for optical frequency standards 
\cite{Ovsiannikov}, that are in turn useful to probe temporal and spatial variations of the fundamental constants \cite{book1} and 
improving global positioning systems \cite{Hong}. Knowing $\lambda_{\rm{magic}}$s of atomic systems are also very useful in the field 
of quantum state engineering \cite{Sackett}, extracting out precise values of the oscillator strengths \cite{tang}, etc. because of
which extensive studies, both experimentally and theoretically, have been carried out in many atoms recently \cite{arora,ab1,
ab2,Hong,Takamoto}. On the other hand, singly charged alkaline earth ions are advantageous to carry out very high precision
measurements using ion trapping and laser cooling techniques. Some of the prominent examples are the optical frequency standards 
\cite{Kajita,Dube}, probing variation of fundamental constants \cite{book1,Margolis}, parity non-conservation effects \cite{Fortson,
Mukherjee} etc. Advantages of these ions owe to their metastable $D$ states that provide longer probe times during interrogation 
of measurements. To reduce systematics in these measurements, state insensitive measurements can be more pertinent that will require 
knowledge of $\lambda_{\rm{magic}}$s in these ions. In fact, $\lambda_{\rm{magic}}$s for the $4S \rightarrow 3D_{{3/2},{5/2}}$ and 
$4S \rightarrow 4P_{{1/2},{3/2}}$ transitions in Ca$^+$ are recently observed \cite{Liu}.

In this Rapid Communication, we report $\lambda_{\rm{magic}}$s for the $nS-nP_{1/2,3/2}$ and $nS - mD_{3/2,5/2}$ transitions 
in the Mg$^+$, Ca$^+$, Sr$^+$ and Ba$^+$ ions, for the principal quantum numbers of the ground state $n$ and of the first excited 
$D$ states $m$ of the respective ions, for the commonly used linearly polarized lights in the experiments by evaluating dynamic dipole 
polarizabilities of these ions accurately.

Dominant contribution to the change in energy of a state $|j_n, m_n \rangle$  with angular momentum $j_n$ and its component $m_n$ 
of an atomic system due to interaction with an external electric field ${\cal E}(\omega)$ of frequency $\omega$ is given by
\begin{equation}
\triangle{E}_{Stark} \approx - \frac{1}{2} \alpha_n(\omega){\cal E}^2(\omega),
\label{seffect}
\end{equation}
where $\alpha_{n}(\omega)$ is the dynamic dipole polarizability of the atomic state $|j_n, m_n \rangle$ and given by
\begin{equation}
\alpha_n  (\omega) = \alpha_{n}^{(0)}(\omega) +  \frac{3m_{j}^2-j_n(j_n+1)}{j_n(2j_n-1)} \alpha_n^{(2)}(\omega),
\label{jmform}
\end{equation}
for the scalar and tensor components as $\alpha_n^{(0)}(\omega)$ and $\alpha_n^{(2)}(\omega)$, respectively. Expressions for these 
quantities in terms of the reduced electric dipole (E1) matrix elements can be found in Refs. \cite{arora,ab1,ab2}. As discussed in our 
previous work \cite{recent}, we estimate $\alpha_n^{(k)}$s (for $k=0,2$) by expressing
\begin{eqnarray}
\alpha_n^{(k)}  &=& \alpha_n^{(k)}(c) + \alpha_n^{(k)}(vc) + \alpha_n^{(k)}(v)
\label{eq26}
\end{eqnarray}
with the notations $c$, $vc$ and $v$ representing the contributions due to the core, core-valence and valence correlation effects, 
respectively. Dominant valence correlations are estimated by calculating electric dipole (E1) matrix elements between many low-lying
intermediate states and referred to as ``Main'' result. This is evaluated using an all order relativistic coupled-cluster method with the 
singles and doubles approximation (SD method) as described in \cite{Blundell,theory}. Other small contributions from the higher excited 
states (referred to as ``Tail'' contribution), core and core-valence correlations are estimated using lower order perturbative methods 
as described in \cite{recent}. To construct the single particle orbitals for the SD method, we have used total 70 B-spline functions 
with a cavity of radius $R=220$ au. However, a few E1 matrix elements involving the $F_{5/2,7/2}$ states in Ba$^+$ are taken from 
\cite{SahooBa} to use as many as E1 matrix elements for more accurate evaluations of the Main contributions to $\alpha_n$s of the $5D$
states of this ion. Estimated static polarizabilities in the above procedure of the considered ground and excited states of the alkaline 
earth metal ions using these matrix elements and experimental energies from national institute of standards and technology (NIST) 
\cite{NIST1} are listed in the Supplemental material along with the individual core, core-valence and Tail contributions. These values 
are also compared with the available precise experimental results and theoretical calculations. We find excellent agreement between our 
results with these values implying that our calculations are precise enough to predict the magic wavelengths in the considered ions 
reliably.
\begin{figure}[t]
\includegraphics[width=8.5cm,height=6.0cm]{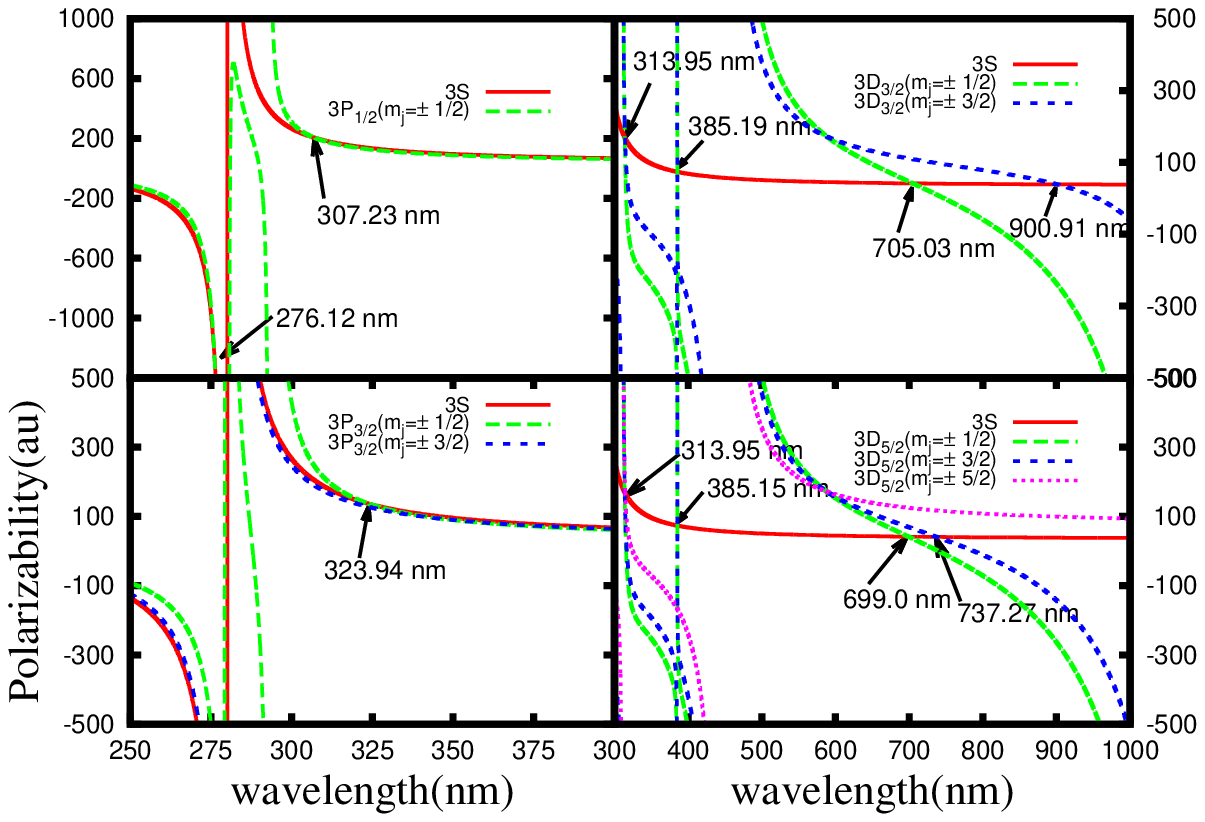}
\caption{(Color online) Dynamic polarizabilities (in au) of the $3S_{1/2}$, $3P_{1/2},_{3/2}$ and $3D_{3/2},_{5/2}$ states in Mg$^+$. Arrows 
indicate positions of magic wavelengths ($\lambda_{\rm{magic}}$s) in nm.} 
\label{Mgsp}     
\end{figure}

\begin{table}[h]{\tiny}
 \caption{\label{magicMg} Magic wavelengths ($\lambda_{\rm{magic}}$s) in nm and corresponding polarizabilities (in au) in the Mg$^+$ ion. 
List of the resonant wavelengths ($\lambda_{\rm{res}}$s) and $\lambda_{\rm{magic}}$s are also given sequentially in increasing order
to indicate their respective placements.}
\begin{ruledtabular}
\begin{tabular}{lcccc}
Resonance & $\lambda_{\rm{res}}$  & $|m_j|$ & $\lambda_{\rm{magic}}$  & $\alpha(\lambda_{\rm{magic}})$\\
\hline 
Transition & \multicolumn{4}{c}{\underline{$3S-3P_{1/2}$}}\\   
$3P_{1/2}$-$3D_{3/2}$ & 279.0777 & 1/2 &  276.12  &  $-1267.54$  \\
$3P_{1/2}$-$3S_{1/2}$ & 280.2704 & \\
$3P_{1/2}$-$4S_{1/2}$ & 292.8683 & 1/2 &  307.23  & 202.59 \\
\hline
Transition & \multicolumn{4}{c}{\underline{$3S-3P_{3/2}$}} \\ 
$3P_{3/2}-3D_{3/2}$ & 279.7930 \\
$3P_{3/2}-3D_{5/2}$ & 279.7998 \\
$3P_{3/2}-3S_{1/2}$ & 279.5528 \\
$3P_{3/2}-4S_{1/2}$ & 293.6510 & 1/2  & 323.94 & 136.51 \\
\hline	    
Transition & \multicolumn{4}{c}{\underline{$3S-3D_{3/2}$}}  \\
$3D_{3/2}-5F_{5/2}$ & 310.4809 & 1/2 & 313.73 &  169.42  \\
	            && 3/2 & 314.18 & 167.53  \\
$3D_{3/2}-5P_{3/2}$ & 384.8335 & 3/2 & 385.01  & 72.56 \\
$3D_{3/2}-5P_{1/2}$ & 385.0385 & 1/2 & 385.38 &  73.47  \\
$3D_{3/2}-4F_{5/2}$ & 448.1327 & 1/2 & 705.03 &  41.44  \\
                    & & 3/2 & 900.91 &  38.67  \\
$3D_{3/2}-4P_{3/2}$ & 1091.5270 & \\
$3D_{3/2}-4P_{1/2}$ & 1095.1779 & \\
\hline
Transition & \multicolumn{4}{c}{\underline{$3S-3D_{5/2}$}} \\ 
$3D_{5/2}-5F_{7/2}$ & 310.4715 & \\
$3D_{5/2}-5F_{5/2}$ & 310.4721 & 1/2 & 313.67 &  169.66  \\
		    && 3/2 & 313.80 & 169.01    \\
		    && 5/2 & 314.39 & 166.70  \\
$3D_{5/2}-5P_{3/2}$ & 384.8209 & 3/2 & 385.12  &73.58 \\
		    && 1/2 & 385.17  &73.56  \\
$3D_{5/2}-4F_{7/2}$ & 448.1130 & \\
$3D_{5/2}-4F_{5/2}$ & 448.1150 & 1/2 & 699.0 &  41.57  \\
		    && 3/2 & 737.27 &  40.79 \\
$3D_{5/2}-4P_{3/2}$ & 1091.423 & \\   
\end{tabular} 
\end{ruledtabular}                   
\end{table}{\tiny}

\begin{figure}[t!]
\includegraphics[width=8.5cm,height=6.0cm]{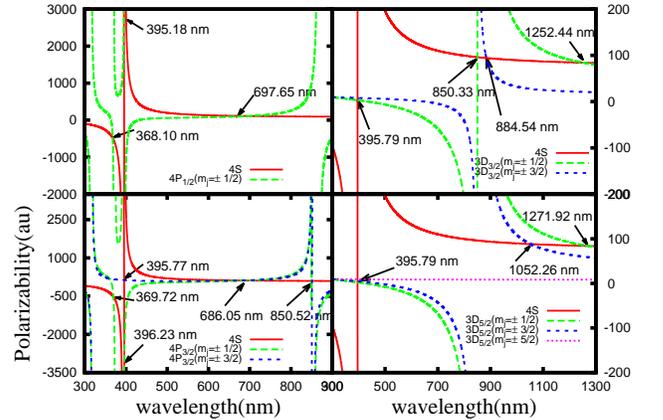}
\caption{(Color online) Dynamic polarizabilities (in au) of the $4S_{1/2}$, $4P_{1/2},_{3/2}$ and $3D_{3/2},_{5/2}$ states 
in Ca$^+$. Locations of average $\lambda_{\rm{magic}}$s (in nm) are indicated by arrows.} 
\label{Casp}     
\end{figure}

\begin{table}[t!]{\tiny}
 \caption{\label{magicCa} $\lambda_{\rm{magic}}$s (in nm) and corresponding polarizabilities (in au) in the Ca$^+$ ion. $\lambda_{\rm{res}}$ 
are also presented to identify the placements of $\lambda_{\rm{magic}}$s clearly 	between two resonant wavelengths. Results 
from this work ([T]) are compared  with the findings from Ref. \cite{tang}.} 
\begin{ruledtabular}
\begin{tabular}{lccccc}
Resonances & $\lambda_{\rm{res}}$ & $|m_j|$ & $\lambda_{\rm{magic}}$ &$\alpha(\lambda_{\rm{magic}})$ & Reference \\
\hline
Transition & \multicolumn{5}{c}{ \underline{$4S-4P_{1/2}$}} \\
\\
$4P_{1/2}-4D_{3/2}$ & 315.887    & 1/2 &  368.10 &  $-485.18$   &  [T] \\
                    &            &     &  368.0149 &  $-477.2554$  & \cite{tang} \\        
$4P_{1/2}-5S_{1/2}$ & 370.603    & 1/2 &  395.18 & 2933.47   & [T] \\
                    &            &     &  395.1807 & 2896.7954 & \cite{tang} \\
$4P_{1/2}-4S_{1/2}$ & 396.847    & 1/2 & 697.65 &  110.39  & [T] \\
                    &            &     & 690.817 & 110.0849 & \cite{tang} \\
$4P_{1/2}-3D_{3/2}$ & 866.214    &     \\   
\hline
Transition & \multicolumn{5}{c}{\underline{$4S-4P_{3/2}$}}\\
$4P_{3/2}-4D_{5/2}$ & 317.933 & \\
$4P_{3/2}-4D_{3/2}$ & 318.128 & 1/2 & 369.72 &  $-520.85$  &  [T]  \\
                    &         &     & 369.6393 & $-512.1655$ & \cite{tang} \\
$4P_{3/2}-5S_{1/2}$ & 373.690 & \\
$4P_{3/2}-4S_{1/2}$ & 393.366 & 3/2 & 395.77  &125.27  & [T] \\
                    &         &     &  395.774 & 123.4906 & \cite{tang} \\ 
		    &          & 1/2 & 396.23 & $-3118.47$  &  [T]  \\
		    &          &     & 396.2315 & $-3077.3881$ & \cite{tang} \\               
$4S_{1/2}-4P_{1/2}$ & 396.847 & 3/2 & 678.35  &113.38  & [T] \\
                    &         &     & 672.508 & 113.0150 & \cite{tang} \\
		    &          & 1/2 & 693.76 &  110.96 & [T] \\
		    &          &     & 687.022 & 110.6606 & \cite{tang} \\
$4P_{3/2}-3D_{3/2}$ & 849.802  & 1/2 & 850.12  & 81.57  & [T] \\
                    &         & 3/2 & 850.92 &  96.09 & [T] \\
$4P_{3/2}-3D_{5/2}$ & 854.209 \\
\hline                   
Transition & \multicolumn{5}{c}{\underline{$4S-3D_{3/2}$}}   \\  
$4S_{1/2}-4P_{3/2}$ & 393.366 & 1/2 & 395.80  &3.13 & [T] \\
                    &         &     & 395.7981 & 1.1711 & \cite{tang}  \\
		    & & 3/2 & 395.79 & 8.02  & [T] \\
		    & &     & 395.7970 & 7.1019 & \cite{tang} \\
$4S_{1/2}-4P_{1/2}$ & 396.847 & \\	
$3D_{3/2}-4P_{3/2}$ & 849.802 & 1/2 & 850.33 & 96.12  & [T] \\
                    &         &     & 850.335 & 95.0011 & \cite{tang} \\
$3D_{3/2}-4P_{1/2}$ & 866.214 & 3/2 & 884.54 & 94.23  & [T] \\
                    &         &     & 887.382 & 92.9908 & \cite{tang} \\
                    && 1/2 & 1252.44 & 84.17 & [T] \\
                    & &   & 1308.590 & 82.4644 & \cite{tang} \\
                    \hline
Transition  & \multicolumn{5}{c}{\underline{$4S-3D_{5/2}$}}   \\
$4S_{1/2}-4P_{3/2}$ & 393.366 & 1/2 & 395.79 & 2.44  & [T] \\
                    &         &     & 395.7982 & 0.5371 & \cite{tang} \\
                     && 3/2 & 395.79& 3.49   & [T] \\
                     & &    & 395.7978 & 3.1792 & \cite{tang} \\
                     && 5/2 & 395.79 & 8.83  & [T] \\
                     & &  &  395.7968 & 8.4633 & \cite{tang} \\
$4S_{1/2}-4P_{1/2}$ & 396.847 & \\
$3D_{5/2}-4P_{3/2}$ & 854.209 & 3/2 & 1052.26 & 88.06 & [T] \\
                    &         &     & 1074.336 & 86.4837 & \cite{tang} \\
                   && 1/2 & 1271.92 & 83.90 & [T] \\ 
                   & &    & 1338.474 & 82.1167 & \cite{tang} \\
\end{tabular} 
\end{ruledtabular}                   
\end{table}{\tiny}

\begin{figure}[t!]
\includegraphics[width=8.5cm,height=6.0cm]{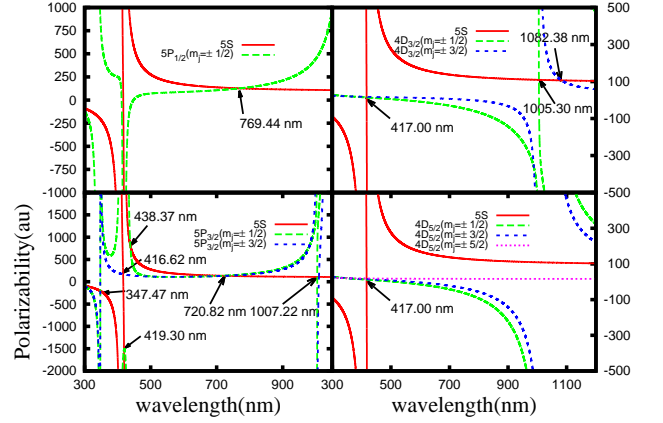}
\caption{(Color online) Dynamic polarizabilities (in au) of the $5S_{1/2}$, $5P_{1/2},_{3/2}$ and $4D_{3/2},_{5/2}$ states
in Sr$^+$. Arrows indicate positions of the average $\lambda_{\rm{magic}}$ in nm.} 
\label{Srsp}     
\end{figure}

\begin{figure}[t!]
\includegraphics[width=8.5cm,height=6.0cm]{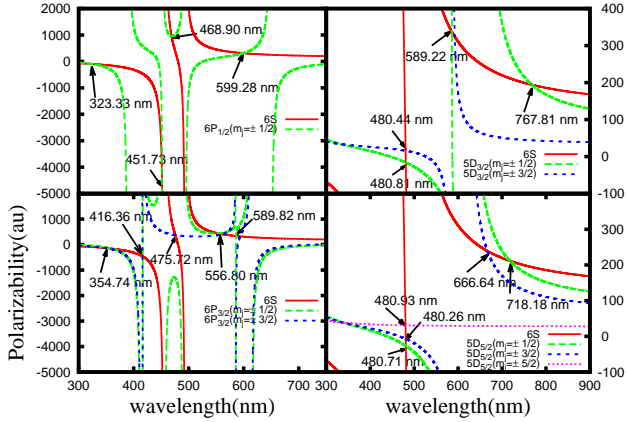}
\caption{(Color online) Dynamic polarizabilities (in au) of the $6S_{1/2}$, $6P_{1/2},_{3/2}$ and $5D_{3/2},_{5/2}$ states
in Ba$^+$ with their average $\lambda_{\rm{magic}}$s (in nm) pointed by arrows.} 
\label{Basp}     
\end{figure}

In Tables \ref{magicMg}, \ref{magicCa}, \ref{magicSr} and \ref{magicBa}, we list $\lambda_{\rm{magic}}$s for the $nS-nP_{1/2,3/2}$ 
and $nS-mD_{3/2,5/2}$ transitions in the Mg$^+$, Ca$^+$, Sr$^+$ and Ba$^+$ ions respectively. They are obtained by locating the crossings 
between the dynamic polarizabilities of the $nS$, $nP_{1/2}$, $nP_{3/2}$, $mD_{3/2}$, and $mD_{5/2}$ states of Mg$^+$,Ca$^+$, Sr$^+$ and 
Ba$^+$ plotted against the frequencies as shown in Figs. \ref{Mgsp}, \ref{Casp}, \ref{Srsp} and \ref{Basp} respectively. The arrows in the 
figures indicate the positions of  $\lambda_{\rm{magic}}$(avg) values which are the average of the magic wavelengths at different $m_j$ 
sub-levels. The $\lambda_{\rm{magic}}$(avg) are not presented for cases where magic wavelengths are (a) not identified for all $m_j$ 
sub-levels, (b) supporting different kind of trapps for all $m_j$ sublevels (c) separated by large wavelengths for different $m_j$ 
sub-levels. We also 
give the total $\alpha_n$ values at the corresponding $\lambda_{\rm{magic}}$s considering different $m_j$ values in the tables. Before we proceed in presenting 
$\lambda_{\rm{magic}}$s, we would like to clarify that $\alpha_n$s of the $nS$ states can have two important resonant transitions ($nS-nP_{1/2}$
and $nS-nP_{3/2}$) in the wavelength range considered in this work. Therefore, $\alpha_{n}$s of the $nS$ states are generally small except 
in the close vicinity of these transitions. In contrast, $\alpha_n$s of the $nP$ and $mD$ states can have significant contributions from several 
resonant transitions in the considered wavelength range. Thus, they are expected to cross with $\alpha_n$s of the $nS$ states in between 
these resonant transitions. We also list the resonant wavelengths ($\lambda_{\rm{res}}$) in the above tables to highlight respective 
placements of $\lambda_{\rm{magic}}$s. We find similar trends in locating magic wavelengths for all the considered ions, except for few 
cases where $\lambda_{\rm{magic}}$s are missing. 

{\bf Mg$^+$:} We are able to locate two $\lambda_{\rm{magic}}$s magic wavelengths for the $3S-3P_{1/2}$ transition in Mg$^+$, one just 
before the $3P_{1/2}-3D$ resonant transition at 276.12 nm and one just after the $3P_{1/2}-4S$ resonant transition at 307.23 nm. The first 
$\lambda_{\rm{magic}}$ supports a blue detuned trap, whereas the second one a red detuned trap. However, only one $\lambda_{\rm{magic}}$ 
for the $3S-3P_{3/2}$($m_j=\pm 1/2$) transition is located at 323.94 nm and nothing is found for the $3S-3P_{3/2}$($m_j=\pm 3/2$) 
transition. On the other hand, three $\lambda_{\rm{magic}}$s are identified for the $3D_{3/2}- 3S$ transition placed between six 
different resonances. First one is located around 314 nm between the $3D_{3/2}-5F_{5/2}$ and $3D_{3/2}-5P_{3/2}$ resonant transitions. The 
next $\lambda_{\rm{magic}}$ is located at the sharp intersection of polarizability curves close to the $3D_{3/2}-5P_{3/2}$ and 
the $3D_{3/2}-5P_{1/2}$ resonances as seen in Fig. \ref{Mgsp}. The last $\lambda_{\rm{magic}}$  is located at 705.03 nm for $m_j=\pm 1/2$ and 
900.91 nm for $m_j= \pm 3/2$ sub-levels, hence it is of limited experimental use. For the $3S-3D_{5/2}$ transition, only one $\lambda_{\rm{magic}}$ appears 
for all $m_j$ sub-levels
around 314 nm. All the observed $\lambda_{\rm{magic}}$s along with the $\lambda_{\rm{res}}$s are 
tabulated in Table \ref{magicMg}. It can be noticed from this table that most of these magic wavelengths favor weak red detuned traps. 

\begin{table}{\tiny}
 \caption{\label{magicSr} $\lambda_{\rm{magic}}$s (in nm) and their corresponding polarizabilities (in au) in Sr$^+$ and $\lambda_{\rm{res}}$ contributing to the polarizabilities of the transitional states are also listed. }
\begin{ruledtabular}
\begin{tabular}{lcccc}
Resonances      & $\lambda_{\rm{res}}$ & $|m_j|$ & $\lambda_{\rm{magic}}$  &  $\alpha(\lambda_{\rm{magic}})$  \\
\hline
Transition & \multicolumn{4}{c}{\underline{ $5S-5P_{1/2}$}} \\
$5P_{1/2}-5D_{3/2}$ & 338.0711 \\
$5P_{1/2}-6S_{1/2}$ & 416.1796 \\
$5P_{1/2}-5S_{1/2}$ & 421.5524  & 1/2 & 769.44 & 126.34  \\
$5P_{1/2}-4D_{3/2}$ & 1091.4874    &         &          \\                    
\hline
Transition & \multicolumn{4}{c}{\underline{ $5S-5P_{3/2}$}} \\
$5P_{3/2}-5D_{5/2}$ & 346.4457 & 3/2  & 347.38 & $-204.86$ \\
$5P_{3/2}-5D_{3/2}$ & 347.4887	  & 1/2 & 347.57  & $-205.66$ \\
$5P_{3/2}-5S_{1/2}$ & 407.7714 & 3/2 & 416.62 & 168.54 \\
	            &           & 1/2 & 419.30 & $-1521.68$ \\
$5S_{1/2}-5P_{1/2}$ & 421.5524 \\
$5P_{3/2}-6S_{1/2}$ & 430.5447 & 1/2 & 438.37 & 817.75 \\
		    &          & 1/2 & 716.72 & 134.25 \\
		    &          & 3/2 & 724.92 & 132.83 \\
$5P_{3/2}-4D_{3/2}$ & 1003.6654  & 1/2 & 1004.47 & 108.08 \\
                    &         & 3/2 & 1009.80 & 108.90  \\
$5P_{3/2}-4D_{5/2}$ & 1032.7309 \\
\hline
Transition & \multicolumn{4}{c}{\underline{$5S-4D_{3/2}$}}   \\ 
$5S_{1/2}-5P_{3/2}$ & 407.7714  & 3/2 & 417.00 & 17.36 \\
		    & & 1/2 & 417.00 & 13.83 \\
$5S_{1/2}-5P_{1/2}$ & 421.5524 \\
$4D_{3/2}-5P_{3/2}$ & 1003.6654 & 1/2 & 1005.30 &108.93 \\
		    && 3/2 & 1082.38  &106.29 \\
$4D_{3/2}-5P_{1/2}$ & 1091.4874 & \\	
\hline
Transition & \multicolumn{4}{c}{\underline{$5S-4D_{5/2}$}}   \\
$5S_{1/2}-5P_{3/2}$ & 407.7714 & 1/2 & 417.01  & 12.16   \\
                     && 5/2 & 417.00 & 18.43 \\
                     && 3/2 & 417.00  &14.25 \\
$5S_{1/2}-5P_{1/2}$ & 421.5524 & \\
$4D_{5/2}-5P_{3/2}$ & 1032.7309 & \\          
\end{tabular} 
 \end{ruledtabular}
\end{table}{\tiny}

\begin{table}[h!]{\tiny}
\caption{\label{magicBa} $\lambda_{\rm{magic}}$s (in nm) and their corresponding polarizabilities (in au) in Ba$^+$.  
$\lambda_{\rm{res}}$s are also given to identify the placements of $\lambda_{\rm{magic}}$s between them.}
\begin{ruledtabular}
\begin{tabular}{lcccc}
Resonances    &$\lambda_{\rm{res}}$ & $|m_{j}|$ & $\lambda_{\rm{magic}}$  & $\alpha(\lambda_{\rm{magic}})$ \\
\hline
Transition & \multicolumn{4}{c}{\underline{$6S-6P_{1/2}$}} \\
$6P_{1/2}-8S_{1/2}$ & 264.72608 & 1/2 &323.33 & $-93.74$  \\ 
$6P_{1/2}-6D_{3/2}$ & 389.17790  & 1/2 &451.73 & $-4574.49$ \\
$6P_{1/2}-7S_{1/2}$ & 452.4926   & 1/2& 468.90 & 953.80 \\
$6P_{1/2}-6S_{1/2}$ & 493.4077   & 1/2 & 599.28 & 311.58  \\
$6P_{1/2}-5D_{3/2}$ & 649.6898 & \\
\hline 
Transition & \multicolumn{4}{c}{\underline{$6S-6P_{3/2}$}} \\ 
$6P_{3/2}-8S_{1/2}$ & 277.13528 & 3/2 & 348.85 & $-133.47$ \\
	            &           & 1/2 & 360.63 & $-158.70$ \\
$6P_{3/2}-6D_{5/2}$ & 413.06491 & \\
$6P_{3/2}-6D_{3/2}$ & 416.60014 & 1/2 & 416.66 & $-460.38$ \\
	            &           & 3/2 & 416.06 & $-459.32$  \\
$6P_{3/2}-6S_{1/2}$ & 455.4033 & 3/2 & 475.72 & 374.67 \\
$6P_{3/2}-7S_{1/2}$ & 489.9927  & 1/2 & 552.60 & 441.01  \\
		    &          & 3/2 & 561.01 & 406.42  \\
$6S_{1/2}-6P_{1/2}$ & 493.4077  & \\
$6P_{3/2}-5D_{3/2}$ & 585.3675  & 1/2 & 586.24 & 336.10  \\
		    &          & 3/2 & 593.40 & 321.95 \\
$6P_{3/2}-5D_{5/2}$ & 614.1713	\\	    
\hline 
Transition & \multicolumn{4}{c}{\underline{$6S-5D_{3/2}$}} \\
 $6S_{1/2}-6P_{3/2}$ & 455.4033 & 3/2 & 480.44 & 15.88 \\
		    & & 1/2 & 480.81 & $-15.82$ \\
$6S_{1/2}-6P_{1/2}$ & 493.4077 &   \\	
$5D_{3/2}-6P_{3/2}$ & 585.3675 & 1/2 & 585.98 & 336.18  \\
		    && 3/2 & 592.46 & 323.72 \\
$5D_{3/2}-6P_{1/2}$ & 649.6898 & 1/2 & 767.81 & 194.05  \\
\hline
Transition& \multicolumn{4}{c}{\underline{$6S-5D_{5/2}$}} \\
$6S_{1/2}-6P_{3/2}$ & 455.4033 & 5/2 & 480.26 & 31.04  \\
                     && 3/2 & 480.71 & $-7.19$ \\
                     && 1/2 & 480.93 & $-26.46$  \\
$6S_{1/2}-6P_{1/2}$ & 493.4077   &\\                    
$5D_{5/2}-6P_{3/2}$ & 614.1713 & 3/2 & 666.64 & 239.13  \\
                   && 1/2 & 718.18 & 211.13  \\
\end{tabular}                   
\end{ruledtabular}
\end{table}{\tiny}
{\bf Ca$^+$:} At least three $\lambda_{\rm{magic}}$s are located for the $4P_{1/2}-4S$ transition in Ca$^+$ at 368.10, 395.18 and 697.65 nms 
between the $4P_{1/2}-4D_{3/2}$, $4P_{1/2}-5S$, $4P_{1/2}-4S$ and $4P_{1/2}-3D_{3/2}$ resonant transitions. A good agreement is found 
between our results and findings by Tang et. al.~\cite{tang} except for our $\lambda_{\rm{magic}}$ at 697.65 nm, which is red detuned as 
compared to their value. Similarly, a total of four $\lambda_{\rm{magic}}$s are located for the $4S-4P_{3/2}$ transition as shown in Fig. \ref{Casp} 
and listed in Table \ref{magicCa}. We also find $\lambda_{\rm{magic}}$s at 395.77 nm and 396.23 nm for the $4S-4P_{3/2}$ transition 
within the fine structure splittings of the $4P$ states which coincides with the experimental observations at 395.7992(7) nm and 395.7990(7)
nm respectively \cite{Liu}. Similar to the $4P_{1/2}-4S$ transition, the other observed $\lambda_{\rm{magic}}$s at 678.35 nm and 693.76 nm are red 
detuned as compared to findings in Ref.~\cite{tang}. Moreover, they have missed a $\lambda_{\rm{magic}}$ around 850 nm which lies 
in between the $4P_{3/2}-3D_{3/2}$ and $4P_{3/2}-3D_{5/2}$ resonant transitions. 
$\lambda_{\rm{magic}}$s for the $4S-3D_{3/2}$ and $4S-3D_{5/2}$ transitions, detected between the fine structure splitting of the 
$4P$ state at 395.79 nm,  offer a scope to trap Ca$^+$ at very small polarizabilities and are not much of practical use. 
As seen in Fig. \ref{Casp}, $\alpha_n$s of the $3D_{5/2}$($m_j=\pm 5/2$) state attain a constant value through out the wavelength 
range considered in this work. Consequently, only one crossing is noticed for this state at 395 nm when $\alpha_n$ of the $4S$ state 
changes sign close to the $4S-4P$ resonances. The other identified $\lambda_{\rm{magic}}$s at 1052.26, 1252.44 and 1271.92 nms for the 
$4S-3D_{3/2}$ and $4S-3D_{5/2}$ transitions with different $m_j$ values are blue detuned compared to the earlier findings of 
Ref.~\cite{tang}. The possible reasons for these disagreements could be that the plots for $\alpha_n$s of the transitional states cross 
at very small angles at these wavelengths thus making them very in-distinctive. Moreover, Tang \textit{et. al.} have performed semi-empirical calculations 
of polarizabilities in the relativistic framework that do not take into account electron correlation effects rigorously like our SD 
method.

{\bf Sr$^+$:}  From Fig. \ref{Srsp}, we are able to locate only one magic wavelength at $769.44$ nm for the $5S$ - $5P_{1/2}$ transition
of Sr$^+$. As given in Table \ref{magicSr}, $\lambda_{\rm{magic}}$s for the $5S-5P_{3/2}$ transition are, however, systematically placed 
between various resonances. We also locate a crossing between the $5S$ and $5P_{3/2}$($m_j=\pm 1/2$) states at 438.37 nm. All identified
$\lambda_{\rm{magic}}$s for the $5S-5P_{3/2}$ transition are typically small in magnitude with exception for $\lambda_{\rm{magic}}=419.302$ 
nm. At this $\lambda_{\rm{magic}}$, it is recommended to use blue tuned trap for the $m_j=\pm 1/2$ sub-levels of Sr$^+$. For the $4D_{3/2} 
\rightarrow 5S$ transition,  $\lambda_{\rm{magic}}=1007.22$ nm lies in the infrared region and is recommended for red detuned trap. In
Fig. \ref{Srsp}, we also observe that $\alpha_n$s for the transitional states at $\lambda_{\rm{magic}}= 417$ nm for the $5S-4D_{3/2}$ 
and $5S-4D_{5/2}$ transitions are very small (almost approach to zero). Therefore, it may not be advantageous to trap the Sr$^+$ ion at 
this wavelength.

{\bf Ba$^+$:} Four $\lambda_{\rm{magic}}$s are found for the $6S-6P_{1/2}$ transition in Ba$^+$ at $323.33$, $451.73$, $468.90$ and $599.28$ 
nms among which $\lambda_{\rm{magic}}=451.73$ nm is in direct vicinity of the $6P_{1/2}-7S_{1/2}$ resonance (within 0.8 nm) yielding a very 
large negative value of $\alpha_n$. However, the other $\lambda_{\rm{magic}}$s are at far distances from the photon excitation energies of 
the resonant transitions yielding small $\alpha_n$s. We also identify at least five $\lambda_{\rm{magic}}$s for the $6P_{3/2}$-$6S_{1/2}$ 
transition placed systematically between various resonant transitions. Similarly, several $\lambda_{\rm{magic}}$s are also located 
for the $6S$-$5D_{3/2}$ and $6S$-$5D_{5/2}$ transitions as seen in Fig. \ref{Basp} in the wavelength range $300$-$800$ nm and listed 
in Table \ref{magicBa}.

{\it Conclusion:}  By determining dynamic dipole polarizabilities of the ground $nS$ and the first excited $nP_{1/2,3/2}$ and $mD_{3/2,5/2}$ 
states, many magic wavelengths of the $nS - nP_{1/2,3/2}$ and $nS - mD_{3/2,5/2}$ transitions of the Mg$^+$, Ca$^+$, 
Sr$^+$ and Ba$^+$ alkaline earth metal ions are identified. Occurrence of magic wavelengths are predicted between the resonant wavelengths 
which will be very helpful to perform high precision measurements in the above ions using both the red and blue tuned trapping techniques.

{\it Acknowledgement:} The work is supported by CSIR Grant No. 03(1268)/13/EMR-II, India, and UGC-BSR Grant No. F.7-273/2009/BSR. 
A part of the computations were carried out using Vikram-100 HPC cluster of Physical Research Laboratory and the employed SD method 
was developed in the group of Professor M. S. Safronova of Delaware University, USA.

\bibliography{magic.bib}

\end{document}


\section*{Supplemental Material}

In a sum-over-states approach the valence correlation contributions can be estimated using the expression
\begin{eqnarray}{\small}
\alpha_n^{(0)}(v) &=& -\frac{2}{3(2j_{n}+1)} \sum_{p \ne n} \frac{ (E_p - E_n) |\langle j_n || D || j_p\rangle |^2} {(E_n-E_p)^2-\omega^2}.
\label{scalar}
\end{eqnarray}{\small}
and
\begin{eqnarray}{\small}
\alpha_n^{(2)}(v) &=& -8  \sum_{p \ne n} (-1)^{j_n+j_p+1}  \left\{\begin{array}{ccc} j_n &  1 & j_n\\ 
	1 & j_p & 2\end{array}\right\} \nonumber \\ && \times \sqrt{\frac{5j_{n}(2j_{n}-1)}{6(j_{n}+1)(2j_{n}+1)(2j_{n}+3)}} \nonumber \\				
  && \times  \frac{ (E_p - E_n) | \langle j_n || D || j_p\rangle |^2 }{(E_n-E_p)^2-\omega^2},
\label{tensor}
\end{eqnarray}{\small}
where $p$ stands for the excited intermediate states, $j$s are angular momentum of the correspond state, $E$s are their energies and 
$D$ is the dipole operator. It is required to calculate a sufficient number of atomic states so that as many as E1 matrix elements 
$\langle j_n || D || j_p \rangle $ can be evaluated to estimate contributions to the $\alpha_n^{(k)}$ values. Contributions from the 
higher excited states, especially from the continuum, are usually small and they can be estimated using a lower order many-body method 
to a reasonable accuracy as suggested in our previous work \cite{recent}. We refer the above described dominant contribution as ``Main''
and the later contribution as ``Tail'' to $\alpha_n^{(k)}(v)$. 

Similarly, the core-valence contribution can be estimated by expressing as
\begin{eqnarray}{\small}
\alpha_n^{(0)}(cv) &=& -\frac{2}{3(2j_{n}+1)} \sum_c  \frac{ (E_c - E_n) |\langle j_n || D || j_c \rangle |^2} {(E_n-E_c)^2-\omega^2}.
\label{scalar}
\end{eqnarray}{\small}
and
\begin{eqnarray}{\small}
\alpha_n^{(2)}(cv) &=& -8  \sum_c (-1)^{j_n+j_c+1}  \left\{\begin{array}{ccc} j_n &  1 & j_n\\ 
	1 & j_c & 2\end{array}\right\} \nonumber \\ && \times \sqrt{\frac{5j_{n}(2j_{n}-1)}{6(j_{n}+1)(2j_{n}+1)(2j_{n}+3)}} \nonumber \\				
  && \times  \frac{ (E_c - E_n) | \langle j_n || D || j_c\rangle |^2 }{(E_n-E_c)^2-\omega^2},
\label{tensor}
\end{eqnarray}{\small}
where $c$ stands for core orbitals. This basically subtracts the extra contributions accounted by allowing the Pauli's exclusion principle 
of the valence electron in the determination of the atomic states. Typically, these contributions are extremely small in magnitude and can be 
estimated using a lower order many-body theory as in Ref. \cite{recent}. 

  Expressions for the core contributions is given by
\begin{eqnarray}{\small}
\alpha_n^{(0)}(c) &=& -\frac{2}{3} \sum_{c,p}  \frac{ (E_c - E_p) |\langle j_p || D || j_c \rangle |^2} {(E_p-E_c)^2-\omega^2}.
\label{scalar}
\end{eqnarray}{\small}
and
\begin{eqnarray}{\small}
\alpha_n^{(2)}(c) &=& -8  \sum_{c,p} (-1)^{j_p+j_c+1}  \left\{\begin{array}{ccc} j_p &  1 & j_p\\ 
	1 & j_c & 2\end{array}\right\} \nonumber \\ && \times \sqrt{\frac{5j_{n}(2j_{n}-1)(2j_n+1)}{6(j_{n}+1)(2j_{n}+3)}} \nonumber \\				
  && \times  \frac{ (E_c - E_p) | \langle j_p || D || j_c\rangle |^2 }{(E_p-E_c)^2-\omega^2},
\label{tensor}
\end{eqnarray}{\small}
where $c$ and $p$ represent sums over core and virtual orbitals, respectively. It can be noticed that there appears an extra phase factor 
in the expression for the tensor polarizability. Since $c$ and $p$ sums over a complete set of $c$ and $p$ orbitals of a closed-core, 
therefore core contribution to the tensor polarizability nullifies. Again, the core contributions to the scalar polarizabilities for the 
states having common core are same since its expression is independent of $j_n$. 
 \begin{table*}
\caption{\label{pol} Calculated values of the static dipole polarizabilities of Ca$^+$, Sr$^+$, Ba$^+$ and Ra$^+$ alkaline earth metal ions. 
These values are compared with the other available theoretical and experimental results. References are given in the square brackets.
The first four low-lying allowed transitions were included in the evaluations of the ``Main'' contributions in the Mg$^+$, Ca$^+$ and Sr$^+$
ions, whereas only the first three low-lying allowed transitions were included in the Ba$^+$ ion. In an exception, the first six low-lying E1 
matrix elements were included in the estimation of ``Main'' contributions in the $3D_{3/2},_{5/2}$ states of the Ca$^+$ ion.}
\begin{ruledtabular}
\begin{tabular}{lccccccccc}
{Mg$^+$} &$\alpha^{(0)}$($3S_{1/2}$)&$\alpha^{(0)}$($3P_{1/2}$)&$\alpha^{(0)}$($3P_{3/2}$)&$\alpha^{(2)}$($3P_{3/2}$)& $\alpha^{(0)}$($3D_{3/2}$) & $\alpha^{(2)}$($3D_{3/2}$)& $\alpha^{(0)}$($3D_{5/2}$) &$\alpha^{(2)}$($3D_{5/2}$)\\
\hline\\
 Main   &  34.50  &   30.89   & 31.18   & 1.32  & 185.35   & $-78.41$   & 184.71 &  $-111.13$  \\
 Core    &  0.48    &  0.48      & 0.48     & -  & 0.48      & -      & 0.48    & -       \\
 Core-valence   & $-1.32\times 10^{-2}$ & 0.0 & 0.0  &- &  $-1.12\times 10^{-3}$  &-&  $-1.70\times 10^{-3}$ &-  \\
 Tail           & $6.66\times 10^{-2}$ & 0.23      & 0.23 & $-0.16$ & 3.46 & $-0.73$  &3.46 & $-1.04$  \\
 Total (Present) & 35.04 & 31.60 & 31.88 & 1.157 & 189.29 & $-79.137$ & 188.65 & $-112.18$ \\
 Others ~\cite{mitroy2}  & 35.01   & 31.598 &  31.598 & 1.163  & 188.6 & $-112.1$ &188.6 & $-112.1$ \\
 Experiment ~\cite{snowt}  & 35.02(4)&  -     & - & - & - & - &- &-\\ 
\hline\\
{Ca$^+$} &$\alpha^{(0)}$($4S_{1/2}$)&$\alpha^{(0)}$($4P_{1/2}$)&$\alpha^{(0)}$($4P_{3/2}$)&$\alpha^{(2)}$($4P_{3/2}$)& $\alpha^{(0)}$($3D_{3/2}$) & $\alpha^{(2)}$($3D_{3/2}$)& $\alpha^{(0)}$($3D_{5/2}$) &$\alpha^{(2)}$($3D_{5/2}$)\\
\hline\\
Main         & 72.80 & $-4.97$  & $-2.96$ & 10.74 & 26.23 & $-16.45$  & 25.98  & $-23.11$  \\
Core         &  3.26  &  3.26   & 3.26   & -   & 3.26   & -     & 3.26    & - \\
Core-valence & $-8.85\times 10^{-2}$ & 0.0  & 0.0 &- & $-7.94\times 10^{-3}$ &-& $-1.20\times 10^{-2}$&-   \\
Tail           & 5.87$\times 10^{-2}$ & 2.52 & 2.52 & $-0.66$ & 3.05 & $-0.62$  & 2.84 & $-0.82$    \\
Total (Present) & 76.03  & 0.82  & 2.82 & 10.08 & 32.31 & $-17.02$ & 32.05 & $-23.92$ \\
Others ~\cite{tang} & 75.28   & $-2.774$    & $-0.931$ & 10.12 & 32.99  & $-17.88$  &  32.81 & $-25.16$ \\
Others \cite{safro} & 76.1(5) & $-0.75(70)$ & 1.02(64) & 10.31(28)  &  32.0(3) & $-17.43(23)$  & 31.8(3) & $-24.51(29)$ \\
Experiment~\cite{edward} & 75.3(4)  & -       &     -   & -        & - & - & - \\ 
\hline\\                    
{Sr$^+$}&  $\alpha^{(0)}$($5S_{1/2}$)&$\alpha^{(0)}$($5P_{1/2}$)&$\alpha^{(0)}$($5P_{3/2}$)&$\alpha^{(2)}$($5P_{3/2}$)& $\alpha^{(0)}$($4D_{3/2}$) & $\alpha^{(2)}$($4D_{3/2}$)& $\alpha^{(0)}$($4D_{5/2}$) &$\alpha^{(2)}$($4D_{5/2}$)\\
\hline\\
Main           &   85.73  & $-39.94$ & $-28.44$ & 11.33    & 53.83 & $-34.25$ & 53.63 & $-46.33$  \\
Core           &   4.98    &  4.98   & 4.98     & -     & 4.98   & -    & 4.98    & -  \\
Core-valence   & $-0.19$  & 0.0      & 0.0 &- & $-1.77 \times 10^{-2}$&- & $-2.79 \times 10^{-2}$  &-      \\
Tail           & 1.97$\times 10^{-2}$ & 3.70 & 2.67 & -0.81 & 4.95 & $-1.00$ & 3.48 & $-1.01$   \\
Total(Present) & 90.54 & $-31.27$  & $-20.79$  & 10.52 & 63.74 & $-35.26$ & 62.08 & $-47.35$ \\
Other \cite{SR}  & 89.88   & $-23.13$   &  $-23.13$ & - & 61.77 & - & 61.77 &- \\
Other \cite{sahoo}     & 88.29(1.0)&  - & -  &- &  61.43(52) & $-35.42(25)$ & 62.87(75) & $-48.83(30)$  \\
Experiment ~\cite{Barklem}  & 93.3(9) & -32.6  & -32.6 & -  & 57.0 & - & 57.0 & - \\  
\hline\\
{Ba$^+$}&$\alpha^{(0)}$($6S_{1/2}$)&$\alpha^{(0)}$($6P_{1/2}$)&$\alpha^{(0)}$($6P_{3/2}$)&$\alpha^{(2)}$($6P_{3/2}$)& $\alpha^{(0)}$($5D_{3/2}$) & $\alpha^{(2)}$($5D_{3/2}$)& $\alpha^{(0)}$($5D_{5/2}$) &$\alpha^{(2)}$($5D_{5/2}$)\\
\hline\\
Main          &  114.19  & 6.85 & 32.00 & 5.87  & 39.71 & $-21.92$  & 42.42 & $-30.37$ \\
Core          &  9.35     & 9.35  & 9.35   & -   & 9.35   & -     & 9.35   & -\\
Core-valence  & $-0.38$ & 0.0 & 0.0 &- & $-2.35\times 10^{-2}$ &- & $-3.87 \times 10^{-2}$ &-   \\
Tail           & 1.66 $\times 10^{-2}$ & 4.26      & 4.18 & $-1.17$ & 4.76 & $-1.00$ & 4.80 & $-1.47$ \\
Total (Present) & 123.18 & 20.46 & 45.53 & 4.70 & 53.80 & $-22.92$  & 56.53 & $-31.83$ \\
Other \cite{sahoo}  &  124.26(1.0) & - & - & - & 48.81(46)  & $-24.62(28)$  & 50.67(58)  & $-30.85(31)$ \\
Experiment ~\cite{snow}& 123.88(5) & -  & - & - &  - &-    & - & - &-\\ 
\end{tabular} 
\end{ruledtabular}
\end{table*}

The first four low-lying allowed transitions are included in the evaluations of the ``Main'' contributions in the Mg$^+$, Ca$^+$ and Sr$^+$ ions, whereas only the first three low-lying allowed transitions are included in the Ba$^+$ ion. However, th efirst six low-lying E1 matrix elements are included in the estimation of ``Main'' contributions in the $3D_{3/2},_{5/2}$ states of the Ca$^+$ ion.
 These states are evaluated using a linearized version of the relativistic coupled-cluster method 
with the singles and doubles approximation (SD method) as described in Refs. \cite{Blundell,theory}. Atomic wave function of the 
$|j_n, m_n\rangle$ state in this method is given by
\begin{eqnarray}
|\Psi_n \rangle &=& \left[1+ \sum_{pa}\rho_{pa} a^\dag_p a_a+ \frac{1}{2} \sum_{pqab} \rho_{pqab}a_p^\dag a_q^\dag a_b a_a\right. \nonumber\\
  &&\left. + \sum_{p \ne n} \rho_{pn} a^\dag_p a_n + \sum_{pqa} a_p^\dag a_q^\dag a_a a_n\right] |\Phi_n\rangle,
  \label{expansion}
\end{eqnarray}
where $|\Phi_n\rangle$ is considered to be the Dirac-Fock (DF) wave function constructed by  
\begin{equation}
 |\Phi_n\rangle =a^\dag_n |\Phi_0\rangle
\end{equation}
for the DF wave function $|\Phi_0\rangle$ of the closed-core. In the above expression, $a^\dag_i$ and $a_i$ represent for the creation and 
annihilation operators. The indices $p,q$ represent for the virtual orbitals and the indices $a,b$ refer to the occupied orbitals.
$\rho$s are the amplitudes of the excitations with the subscripts for the corresponding orbitals. The matrix element of $D$ between 
states $|j_vm_v\rangle$ and $j_wm_w\rangle$ states is evaluated in this method by \cite{marianat}.
\begin{equation}
 D_{vw} = \frac{\langle \psi_v|D|\Psi_w\rangle}{\sqrt{\langle\psi_v|\psi_v\rangle\langle\psi_w|\psi_w\rangle}}.
\end{equation}
Combining E1 matrix elements obtained using the SD method with the experimental energies from the national institute of standards and 
technology (NIST) database \cite{NIST1}, we determine the ``Main'' contributions to valence correlations of $\alpha_n$s of the 
$nS_{1/2}$, $nP_{1/2},_{3/2}$ and $mD_{3/2},_{5/2}$ states in the considered alkaline earth metal ions. In Table \ref{pol}, we list the 
contributions from the ``Main'', core, valence-core and ``Tail'' components to the static values of $\alpha_n$ for all the considered 
alkaline ions. The final values are compared with other available theoretical and experimental values. Transitions up to the $3S-6P$, $3P-6S$, $3P-6d$, $3D-6P$ and $3D-7F$ are included into the "Main" contribution for Mg$^+$ ion. As seen from Table \ref{pol}, the 
ground state dipole polarizability values obtained for the Mg$^+$ ion are very close to the values estimated by Mitroy and Zhang 
\cite{mitroy2}. They evaluate the non-relativistic values of polarizabilities by diagonalizing the semi-empirical Hamiltonian in a large 
dimension single electron basis. We notice that there is a large difference between our polarizability results for $3D_{3/2}$ and 
$3D_{5/2}$ states and hence the non-relativistic values calculated by them are unfavorable. 
Snow \cite{snowt} deduced the static polarizability for the ground state of the Mg$^+$ ion from the fine structure of high 
Rydberg levels. Their $\alpha^{\rm{(0)}}$ value of 35.02(0.04) $a_0^3$ agrees well with our result of 35.04 $a_0^3$. 
 For Ca$^+$ ion transitions up to the $4S-7P$, $4P-7S$, $4P-6d$, $3D-7P$ and $3D-7F$ are included into the "Main" contribution. While examining 
polarizabilities in case of the Ca$^+$ ion, we notice that our calculation for the ground state polarizability is in the agreement with the 
result obtained from experimental spectral analysis by Chang \cite{edward}. Also, a comparison of our calculated polarizabilities for this 
ion with the calculations done using combination of many-body perturbation theory and SD scaled method by Safronova \textit{et al.} \cite{safro} 
shows very good agreement. Moreover, our result for the ground state is in consistent agreement with the results calculated by Tang and 
co-workers \cite{tang}. However, a discrepancy has been observed between our results and other calculations among the $P$ state polarizabilities 
this ion. We notice that the polarizabilities of the $P$ states are very small due to substantial cancellations between some large contributions 
to the total polarizability. The final values are, thus, very sensitive to these cancellations. Next, we compare our polarizability results for 
the Sr$^+$ ion in Table \ref{pol}. Transitions up to the $5S-8P$, $5P-8S$, $5P-7d$, $4D-8P$ and $3D-7F$ are included into the "Main" contribution for this ion. Mitroy and coworkers \cite{SR} have used a non-relativistic method using a sum-over-states approach to determine 
polarizabilities of the $S$, $P$ and $D$ states of Sr$^+$. It is evident from Table \ref{pol} that our ground state dipole polarizability is in 
agreement with their result. However, it may not be proper to compare their non-relativistic values of the polarizabilities for the $4D_{3/2}$ 
and $4D_{5/2}$ states just by including corrections from the relativistic effects with the non-relativistic values agnaist our fully relativistic 
calculations. However, the static dipole polarizabilities calculated for the ground and $4D$ state of the Sr$^+$ ion using an {\it ab initio} 
relativistic coupled-cluster method \cite{sahoo} are in close agreement with our results. There are no direct experimental values available to 
compare with these results. The polarizability values for the $S$, $P$ and $D$ states of the Sr$^+$ ion, derived by combining the experimental 
results with oscillator strength sums by Barklem and O'Mara \cite{Barklem}, has a considerable discrepancy with our present results. In 
Ba$^+$ ion, transitions up to the $6S-8P$, $6P-8S$, $6P-6d$, $5D-8P$ and $5D-7F$ are included into the "Main" contribution. The high-precision ground state polarizability measurement for Ba$^+$, achieved by using a novel technique, based on the resonant 
excitation Stark ionization spectroscopy \cite{snow}, is in very good agreement with our work. To summarize, the above analysis clearly justifies that 
our calculations of dipole polarizabilities in the considered alkaline earth metal ions are accurate enough to predict the magic wavelengths in these 
ions reliably.